\documentclass[12pt]{elsart}

\usepackage{graphicx}
\usepackage{cite}
\usepackage{endfloat}

\journal{Journal of Magnetism and Magnetic Materials}

\begin{document}

\begin{frontmatter}

\title{Critical and compensation phenomena in a mixed-spin ternary alloy: a Monte Carlo study}
\author{M. \v{Z}ukovi\v{c}\corauthref{cor}},
\ead{milan.zukovic@upjs.sk}
\author{A. Bob\'ak}
\ead{andrej.bobak@upjs.sk}
\address{Department of Theoretical Physics and Astrophysics, Faculty of Science, P. J. \v{S}af\'arik University, Park Angelinum 9, 041 54 Ko\v{s}ice, Slovak Republic}
\corauth[cor]{Corresponding author.}

\begin{abstract}
By means of standard and histogram Monte Carlo simulations, we investigate the critical and compensation behaviour of a ternary mixed spin alloy of the type $AB_pC_{1-p}$ on a cubic lattice. We focus on the case with the parameters corresponding to the Prussian blue analog $( {\rm Ni}^{\rm II}_p {\rm Mn}^{\rm II}_{1-p})_{1.5}[{\rm Cr^{III}(CN)}_6] \cdot n {\rm H_2O}$ and confront our findings with those obtained by some approximative approaches and the experiments.
\end{abstract}


\end{frontmatter}

\section{Introduction}

Prussian blue analogs, a class of molecular-based magnetic materials, are of high interest due to a relatively easy control of their magnetic properties during a synthesis process through the selection of desired spin sources. These compounds exhibit some remarkable properties (see, e.g.,~\cite{Verd1,Hash1} and references therein), such as a photoinduced magnetic pole inversion, a compensation or multicompensation behaviour, or an anomalous thermal dependence of the susceptibility as the compensation temperature approaches the critical one. In particular, an example of the compensation behaviour is the Prussian blue analog of the type $({\rm Ni}^{{\rm II}}_p{\rm Mn}^{{\rm II}}_{1-p})_{1.5}$ $[{\rm Cr}^{{\rm III}}({\rm C}{\rm N})_6]\cdot n {\rm H}_2 {\rm O}$, which includes both ferromagnetic ($J_{NiCr}>0$) and antiferromagnetic~($J_{MnCr}<0$) superexchange interactions through the cyanide bridging ligands. This so called ternary metal compound, besides a critical temperature, exhibits a compensation temperature for some range of the mixing ratio $p$ of the metal ions, and an unusual behaviour of the susceptibility as the two temperatures merge~\cite{Hash1,Ohko3}. Thus, the compensation behaviour results from a proper selection of the parameter $p$, and its actual value also quantitatively affects the magnitudes of the critical and compensation temperatures. We note that the compensation behaviour, with important technological applications in the field of magneto-optic recording, is possible due to different temperature dependencies of the sublattice magnetizations.

Due to their structural complexity, the theoretical investigations of the Prussian blue analogs have been mostly limited to a mean-field theory (MFT) treatment \cite{Ohko3,Ohko1,Ohko4,Boba1,Dely1} or an effective-field theory (EFT) approach \cite{Boba2,Hu1}, but also Monte Carlo (MC) simulations \cite{Buen1,Tsuj1,Dely2} and exact recursion relations on the Bethe lattice \cite{Devi1} have recently been employed. However, in the previous MC studies only simplified models of the Prussian blue analogs, consisting of a fixed spatial configuration of the spins placed on two equivalent ($1:1$ stoichiometry) interpenetrating sublattices, were considered. Furthermore, either lattice geometry (\cite{Buen1}), spin values (\cite{Dely2}), or both (\cite{Tsuj1}) did not correspond to the compound of our current interest. Therefore, the main goal of the present paper is to focus on a spin system corresponding to the Prussian blue analog $( {\rm Ni}^{\rm II}_p {\rm Mn}^{\rm II}_{1-p})_{1.5}[{\rm Cr^{III}(CN)}_6] \cdot n {\rm H_2O}$, which can be modeled by a three-dimensional $AB_pC_{1-p}$ ternary alloy on a cubic lattice with a $3:2$ stoichiometry, consisting of three different Ising spins $S_A = \frac{3}{2}$, $S_B = 1$, and $S_C = \frac{5}{2}$ and the superexchange interaction ratio $R = |J_{AC}|/J_{AB} \equiv |J_{MnCr}|/J_{NiCr} \approx 0.45$ \cite{Ohko3,Ohko1}. We also take into account fluctuations resulting from different spatial distributions of magnetic and non-magnetic sites. Further, we study how magnetic properties of the ternary metal compound are modified when the parameter $R$ and the stoichiometry of the system are changed.


\section{Model and Monte Carlo simulations}
The mixed ferro-ferrimagnetic ternary alloy model of the type $AB_pC_{1-p}$ is described by the Hamiltonian 
\begin{eqnarray}
\label{Hamil}
H = -\sum_{(i,j)}\xi_iS_i^A[J_{AB}\xi_jS_j^B + J_{AC}S_j^C(1 - \xi_j)],
\end{eqnarray}
where $S_i^A = \pm \frac{3}{2},$ $\pm \frac{1}{2}$ for $A$ ions, $S_j^B = \pm 1, 0$ for $B$ ions, and $S_j^C = \pm \frac{5}{2}$, $\pm \frac{3}{2}$, $\pm \frac{1}{2}$ for $C$~ions. $\xi_i$~is a random variable which takes the value of unity (zero) if the site $i$ is occupied by the $A$ ion (vacant), and $\xi_j$~is a random variable which takes the value of unity (zero) if the site $j$ is occupied by the $B$ $(C)$ ion. To be consistent with the structure of the above-mentioned Prussian blue analog we also assume that the couplings between nearest neighbours include both the ferromagnetic ($J_{AB} > 0$) and the antiferromagnetic ($J_{AC} < 0$) interactions. A simulated $L \times L \times L$ lattice consists of two interpenetrating face-centered cubic sublattices, each one comprising $L^3/2$ sites. The $A$ ions, which are randomly distributed on one sublattice with the concentration $p_A=2/3$, are alternately connected with the $B$ or $C$ ions, which are randomly located on the other sublattice with the concentration $p$ or $1-p$, respectively. Thus, each $A$ ion is coupled with six nearest-neighbouring ions of the type $X$ ($X = B$ or $C$), while the ions of the type $X$ have on average only four nearest neighbours of the type $A$. This arrangement reflects the 3:2 stoichiometry of the system. It should be stressed that in the previous MC studies \cite{Buen1,Tsuj1,Dely2,Dely3} the real structure of the Prussian blue analogs was not taken into account.
 
We consider linear lattice sizes $L$ ranging from $12$ up to $50$ with the periodic boundary conditions imposed. Initial spin states are randomly assigned and the updating follows the Metropolis dynamics. The lattice structure and the short range of
the interactions enable vectorization of the algorithm. Since the spins on one sublattice interact only with the spins on the other, each sublattice can be updated simultaneously. Thus one sweep through the entire lattice involves just two sublattice updating steps. For thermal averaging, we consider up to $1 \times 10^5$ MC sweeps in the standard and up to $1 \times 10^6$ MC sweeps in the histogram MC simulations, discarding the first $20$\% of the total number for thermalization. Finally, configurational averaging is performed over $Z$ (typically 10 or 20) spatial configurations. Then the errors of the calculated quantities are determined from the values obtained for those configurations. Namely, the error bars in all the figures presented below will represent twice of the standard deviations.

We calculate the sublattice magnetization per site
\begin{equation}
\label{Magn_A}
m_A\equiv [\langle M_A \rangle] = \Bigg[\frac{2}{L^3} \Bigg\langle \Bigg|\sum_{i=1}^{N_A}S_{i}^{A}\Bigg| \Bigg\rangle\Bigg],
\end{equation}

\begin{equation}
\label{Magn_B}
m_B\equiv [\langle M_B \rangle] = \Bigg[\frac{2}{L^3} \Bigg\langle \Bigg|\sum_{j=1}^{N_B}S_{j}^{B}\Bigg| \Bigg\rangle\Bigg],
\end{equation}

\begin{equation}
\label{Magn_C}
m_C\equiv [\langle M_C \rangle] = \Bigg[\frac{2}{L^3} \Bigg\langle -\Bigg|\sum_{j=1}^{N_C}S_{j}^{C}\Bigg| \Bigg\rangle\Bigg],
\end{equation}

\noindent where $N_A$ is the number of $A$ ions ($N_A=p_AL^3/2$) on one sublattice and $N_B$, $N_C$ are the numbers of $B$ and $C$ ions ($N_B=pL^3/2$, $N_C=(1-p)L^3/2$) on the other sublattice, and $\langle\cdots\rangle$ denotes thermal and $[\cdots]$ configurational averages. The total magnetization per site is defined as
\begin{equation}
\label{Magn_dir}
m\equiv [\langle M \rangle]=\bigg[\bigg\langle\frac{1}{2}(M_A+M_B+M_C)\bigg\rangle\bigg]= \frac{1}{2}(m_A+m_B+m_C)
\end{equation}
and the staggered magnetization per site is given by
\begin{equation}
\label{Magn_stag}
m_s\equiv [\langle M_s \rangle]=\Bigg[\frac{1}{L^3}\Bigg\langle\Bigg|\sum_{(i,j)}(S_i^A+S_j^B-S_j^C)\Bigg|\Bigg\rangle\Bigg].
\end{equation}
The corresponding direct and staggered susceptibilities per site $\chi_M$ and $\chi_{M_s}$, respectively, are given by
\begin{equation}
\label{Susc_dir}
\chi_{M}=\Bigg[\frac{L^3}{k_{B}T}(\langle M^{2} \rangle - \langle M \rangle^{2})\Bigg],
\end{equation}
and
\begin{equation}
\label{Susc_stag}
\chi_{M_s}=\Bigg[\frac{L^3}{k_{B}T}(\langle M_{s}^{2} \rangle - \langle M_{s} \rangle^{2})\Bigg].
\end{equation}
Finally, the specific heat per site $c$ is defined by
\begin{equation}
\label{Spec_heat}
c=\Bigg[\frac{(\langle H^{2} \rangle - \langle H \rangle^{2})}{L^3k_{B}T^{2}}\Bigg].
\end{equation}

In the histogram MC simulations, developed by Ferrenberg and Swendsen \cite{Ferr1,Ferr2}, we perform longer runs at a chosen temperature and calculate the energy histogram $P(H)$, the order parameters histograms $P(O)$\ $(O=M\ \rm{or}\ M_s)$, as well as the physical quantities (\ref{Magn_A})$-$(\ref{Spec_heat}). Using data from the histograms one can calculate physical quantities at neighbouring temperatures, and thus determine the values of extrema of various quantities and their locations with high
precision for each lattice size. In such a way we can obtain quality data for finite-size scaling (FSS) analysis. The extrema of the direct and staggered susceptibilities at a second-order transition are known to scale with the lattice size as
\begin{equation}
\label{eq.scalchi}
\chi_{O,max}(L) \propto L^{\gamma_{O}/\nu_{O}}\ ,
\end{equation}
\noindent where $\nu_O$, $\gamma_O$ represent the correlation length and susceptibility critical
exponents, respectively, pertaining to the order parameter $O\ (O=M\ \rm{or}\ M_s)$.


\section{Results and discussion}
\subsection{Critical temperature}

\begin{figure}[ht]
\begin{center}
\includegraphics[scale=0.45]{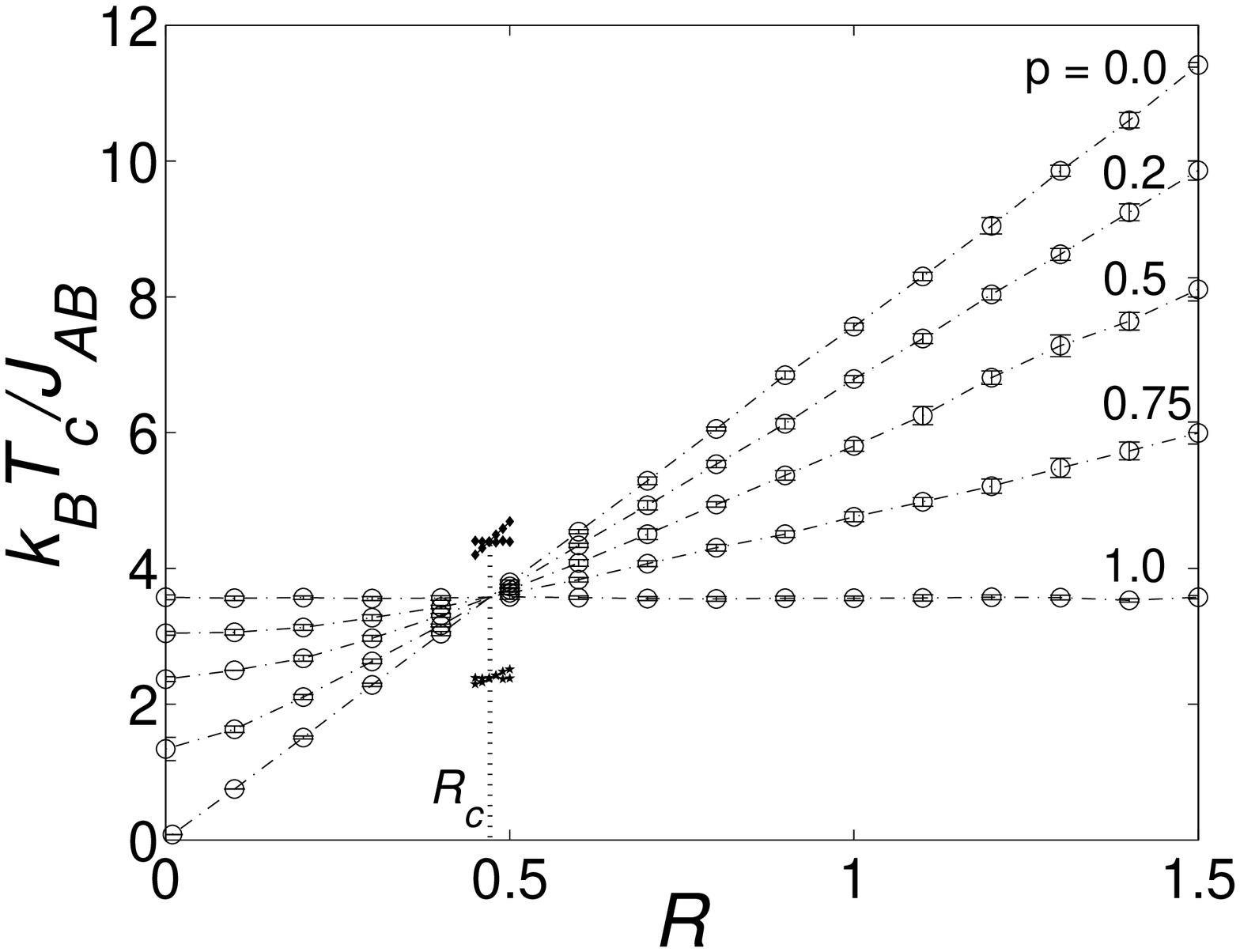}%
\caption{Critical temperature versus interaction ratio $R$ for selected values of the concentration $p$. The dash-doted lines are guides for the eye and the dotted line marks an approximate location of $R_c \approx 0.47$. Above and below the crossing point, the crossing points for the systems with $1:1$ and $3:1$ stoichiometries, respectively, are marked.}
\label{fig:Tc-R}
\end{center}
\end{figure}

The critical temperature of the current model can be considerably affected by the values of the exchange ratio $R$ and the concentration $p$. Phase diagrams in a wide range of the $(p-R)$ parameter space are estimated based on monitoring of the positions of the (critical) maxima in different thermodynamic quantities, such as $\chi_M$, $\chi_{M_s}$, and $c$. Due to the extensive coverage it would be computationally too demanding to perform the FSS analysis in each considered point of the $(p-R)$ plane to determine the critical temperatures with high precision. Therefore, we estimate the values of $T_c$ from the peaks positions of $\chi_{M_s}$ for a fixed value of $L$, i.e., we determine the pseudocritical temperature $T_c(L)$ . After checking in a few selected points we chose the value of $L=20$ above which the values of $T_c(L)$ do not tend to change significantly. In Fig. \ref{fig:Tc-R}, we show dependence of such estimated critical temperature on $R$ for some selected values of $p$.
The limiting cases of $p=0$ and $p=1$ correspond to the binary mixed systems $AC$ with spins $\frac{3}{2}$ and $\frac{5}{2}$, and $AB$ with spins $\frac{3}{2}$ and $1$, respectively. While, the latter binary system is not affected by the parameter $R$ and gives a flat line, the former one gives the transition line with the steepest gradient. Further, from the figure it can be seen that all the curves appear to intersect in a single point corresponding to the critical value $R_c$. By closer inspection of the intersection area for various lattice sizes we locate $R_c \approx 0.47$, a value surprisingly close to that obtained by MFT, $R_c^{MFT} = ( \frac{8}{35} )^{1/2} \approx 0.478$ \cite{Boba1}. This finding comes surprising particularly in the light of the fact that the $R_c^{MFT}$ value does not depend on the coordination numbers (numbers of nearest neighbours) in the respective sublattices, while the current approach accounts for the system's 3:2 stoichiometry. This prompted us to check the value of $R_c$ in our system with different stoichiometries. In the considered cases of $1:1$ ($p_A=1$), $3:2$ ($p_A=2/3$), and $3:1$ ($p_A=1/3$) stoichiometries we find that the critical temperatures decrease with decreasing $p_A$ (hence, decreasing overall coordination number), however, the respective values of $R_c$ are indistinguishable within the current precision. Moreover, they are very close to the value obtained by the MC simulations on a two-dimensional system \cite{Buen1}. These results indicate very little or no sensitivity of the value of $R_c$ to the lattice stoichiometry, supporting the MFT results. However, the value of the corresponding critical temperature obtained by the MC approach is considerably lower than that obtained by MFT. For example, in the system with the $3:2$ stoichiometry the value of the critical temperature from the MC calculation is $k_BT_c^{MC}/J_{AB} \approx 3.57$, while MFT gives $k_BT_c^{MFT}/J_{AB} \approx 4.47$. 

The little sensitivity of the value of $R_c$ to the lattice stoichiometry can be understood considering the processes that the respective parameters control. The parameter $R=|J_{AC}|/J_{AB}$ controls the relative strength of the exchange interactions between $A$ ions and their neighbours on $C$ and $B$ sublattices. At $R_c$, a kind of a balance sets in at which a mutual exchange of $B$ and $C$ ions does not affect the critical temperature of the system. On the other hand, the lattice stoichiometry (controlled by the parameter $p_A$) relates solely to the concentration of $A$ ions. Therefore, while the stoichiometry affects the overall coordination number and hence the magnitude of the critical temperature, it is not likely to affect the relative strength of the exchange interactions between the $AB$ and $AC$ sublattice pairs, i.e. the value of $R_c$. 

\begin{figure}[ht]
\begin{center}
\includegraphics[scale=0.45]{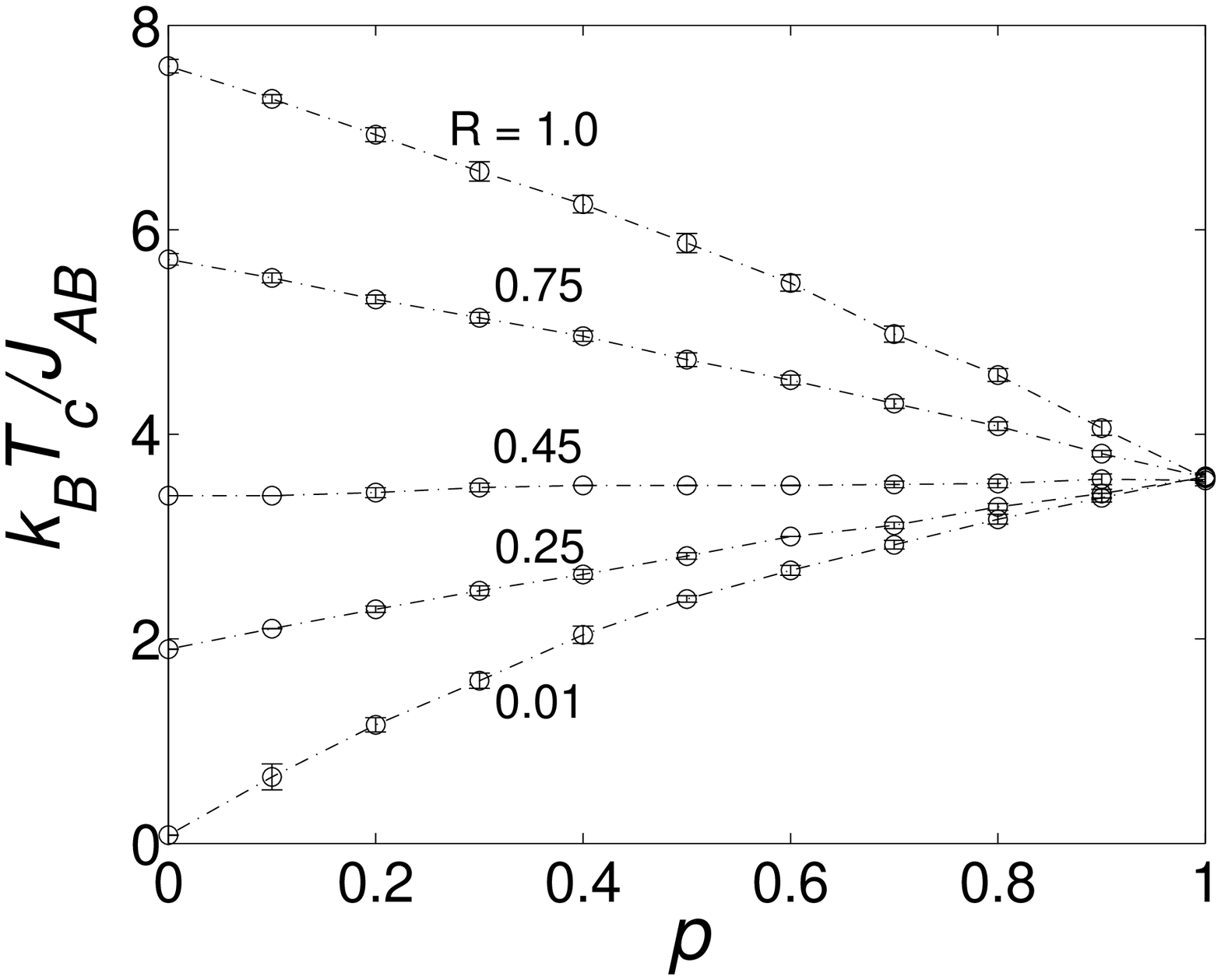}
\caption{Critical temperature as a function of $p$ for different values of $R$.}
\label{fig:Tc-p}
\end{center}
\end{figure}

As we can observe from Fig. \ref{fig:Tc-R}, the transition temperature $T_c$ for the ternary systems with $R<R_c$ is always lower than that of the binary mixture $AB$ and increases with increasing concentration $p$, while the opposite trend is observed for $R>R_c$. This behaviour of $T_c$ is evident from Fig. \ref{fig:Tc-p},
in which the transition temperature is shown as a function of a varying concentration. The transition line with $R=0.45$ corresponds to the real compound $({\rm Ni}^{{\rm II}}_p{\rm Mn}^{{\rm II}}_{1-p})_{1.5}$ $[{\rm Cr}^{{\rm III}}({\rm C}{\rm N})_6]\cdot n {\rm H}_2 {\rm O}$ \cite{Ohko1} and, as the Fig. shows, due to its proximity to the value of $R_c$, the critical temperature is little affected by the change of the mixing ratio $p$. The transition curves merge at $p=1$, for which the ternary system becomes the binary mixture $AB$, and therefore, $R$ becomes a redundant variable. 

\begin{figure}[ht]
\begin{center}
\includegraphics[scale=0.45]{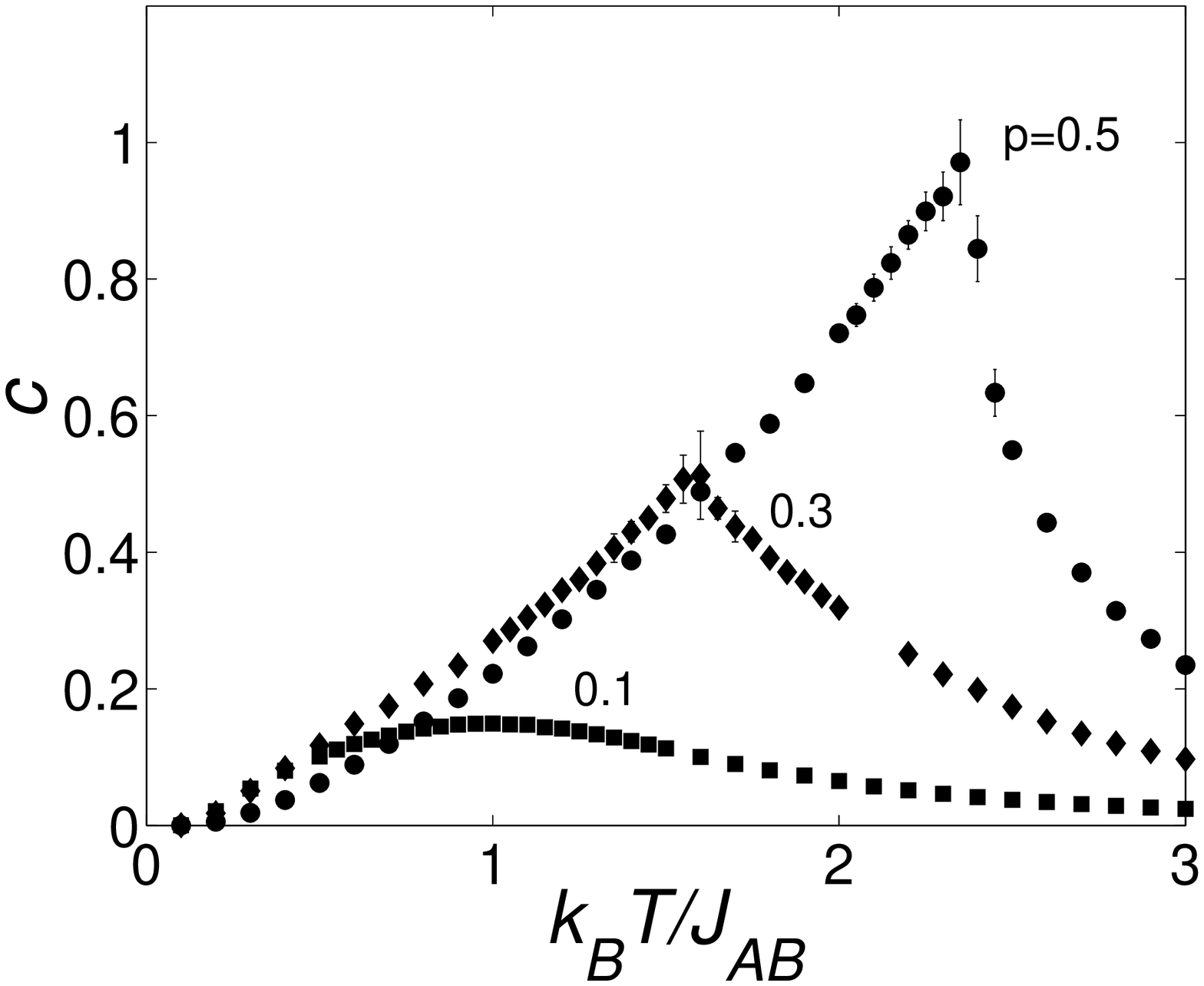}
\caption{Temperature dependences of the specific heat for $R=0$, $L=32$, and different values of $p$.}
\label{fig:c-T}
\end{center}
\end{figure}

An interesting situation arises at the low-concentration side for the vanishing value of $R$. Namely, the transition temperature decreases with the decrease of $R$ but as long as $R$ remains finite the phase transition occurs in the whole concentration range. The situation for $R=0$ corresponds to configurations involving ''loose'' ions of the types $C$ and $A$ with no mutual exchange interactions and their concentration increases with decreasing value of $p$. Therefore, even though we do not deal with a problem of a dilution with non-magnetic impurities, the system involves a portion of spins that in the absence of the exchange interactions behave like non-magnetic atoms, since they are inactive in producing magnetic ordering. For this reason we can expect some ``percolation threshold'' $p_c$ below which no phase transition can occur. Unfortunately, within the current approach it is computationally very demanding to locate the value of $p_c$ with a higher precision. The main reasons are extremely long thermalization times near $p_c$, necessity to consider large lattice sizes, and to average over a large number of configurations. For instance, in our tests on individual configurations for $p=0.2$, the equilibrium was not reached even after $\approx 10^6$ MC sweeps. Nevertheless, from the observation of some physical quantities, as well as the finite-size scaling, it appears that the long-range ordering survives down to at least $p=0.3$. In Fig. \ref{fig:c-T} we show temperature dependence of the specific heat for $p=0.5$, $0.3$ and $0.1$, for the maximum linear size of $L=32$, in order to suppress as much as possible the finite-size effects. The specific heat was chosen due to the fact that it exhibits much smaller fluctuations, and hence ``smoother'' behaviour, than the susceptibility. While the curves for $p=0.5$ and $0.3$ in Fig. \ref{fig:c-T} display well-defined sharp maxima, typical for a second-order phase transition, only a broad maximum is observed for $p=0.1$. 
\begin{figure}[ht]
\begin{center}
\includegraphics[scale=0.45]{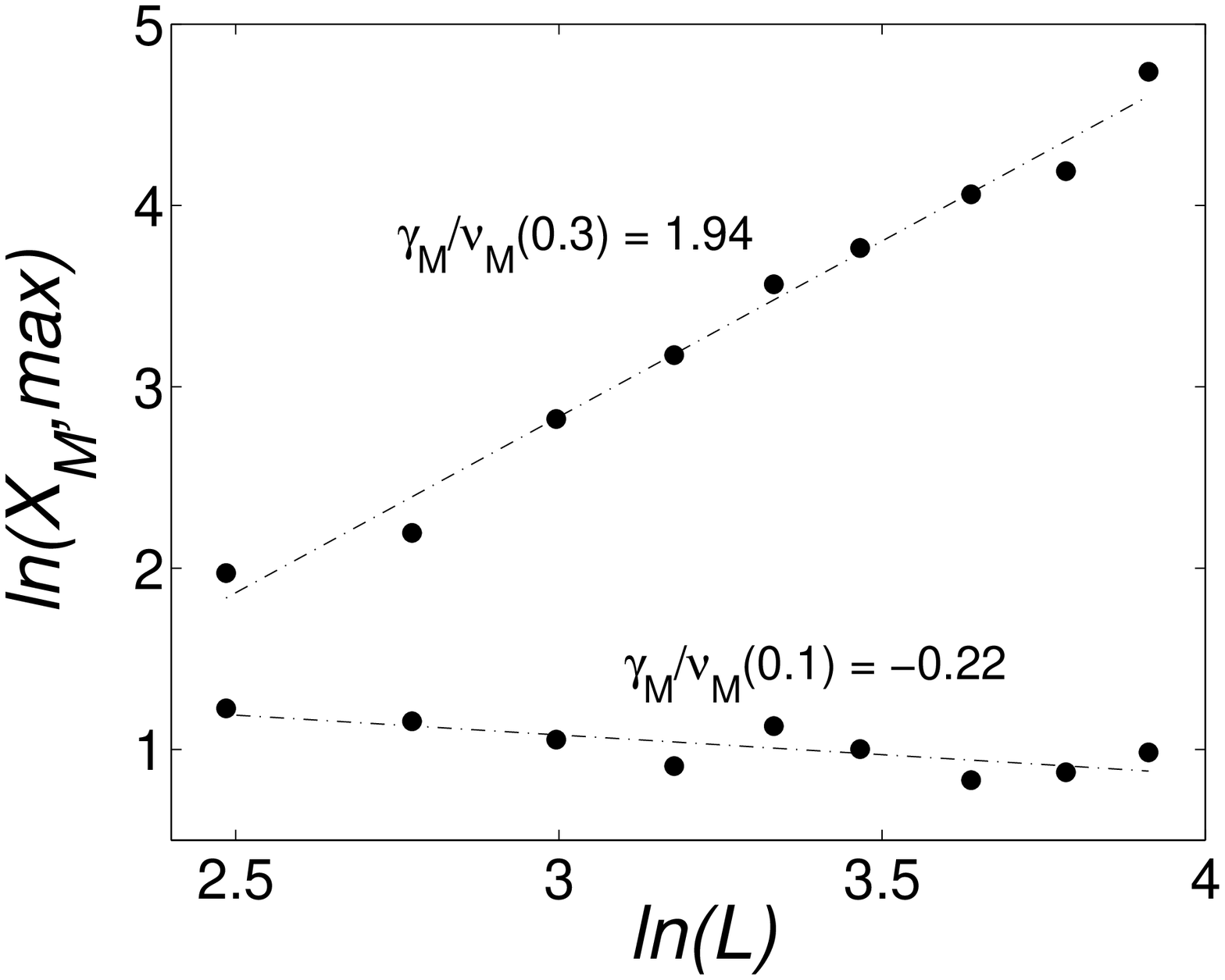}
\caption{Finite-size scaling of the maxima of the direct susceptibility, for $p=0.3$ and $p=0.1$.}
\label{fig:log-log}
\end{center}
\end{figure}
To see qualitative difference between the anomalies in the physical quantities for $p=0.3$ and $0.1$, we additionally performed FSS analysis of the direct susceptibility $\chi_M$ (for $R=0$, there is no reason to consider $\chi_{M_s}$) using the histogram MC simulation. As depicted in Fig. \ref{fig:log-log}, for $p=0.3$ the standard Ising value of the critical exponents ratio $\gamma/\nu \approx 1.97$ \cite{Ferr3} was recovered with a fair precision, while for $p=0.1$ even a negative value was obtained. We conjecture that the anomalies in the response functions for $p=0.1$ originate from the presence of finite clusters of ordered spins rather than the long-range ordering, and therefore, we roughly estimate the value of $p_c$ to fall into the region of $0.1<p_c<0.3$. On the other hand, MFT is not able to deal correctly with the percolation problem and predicts $p_c=0$ \cite{Boba1}.

\subsection{Compensation temperature}

\begin{figure}[b]
\begin{center}
\includegraphics[scale=0.45]{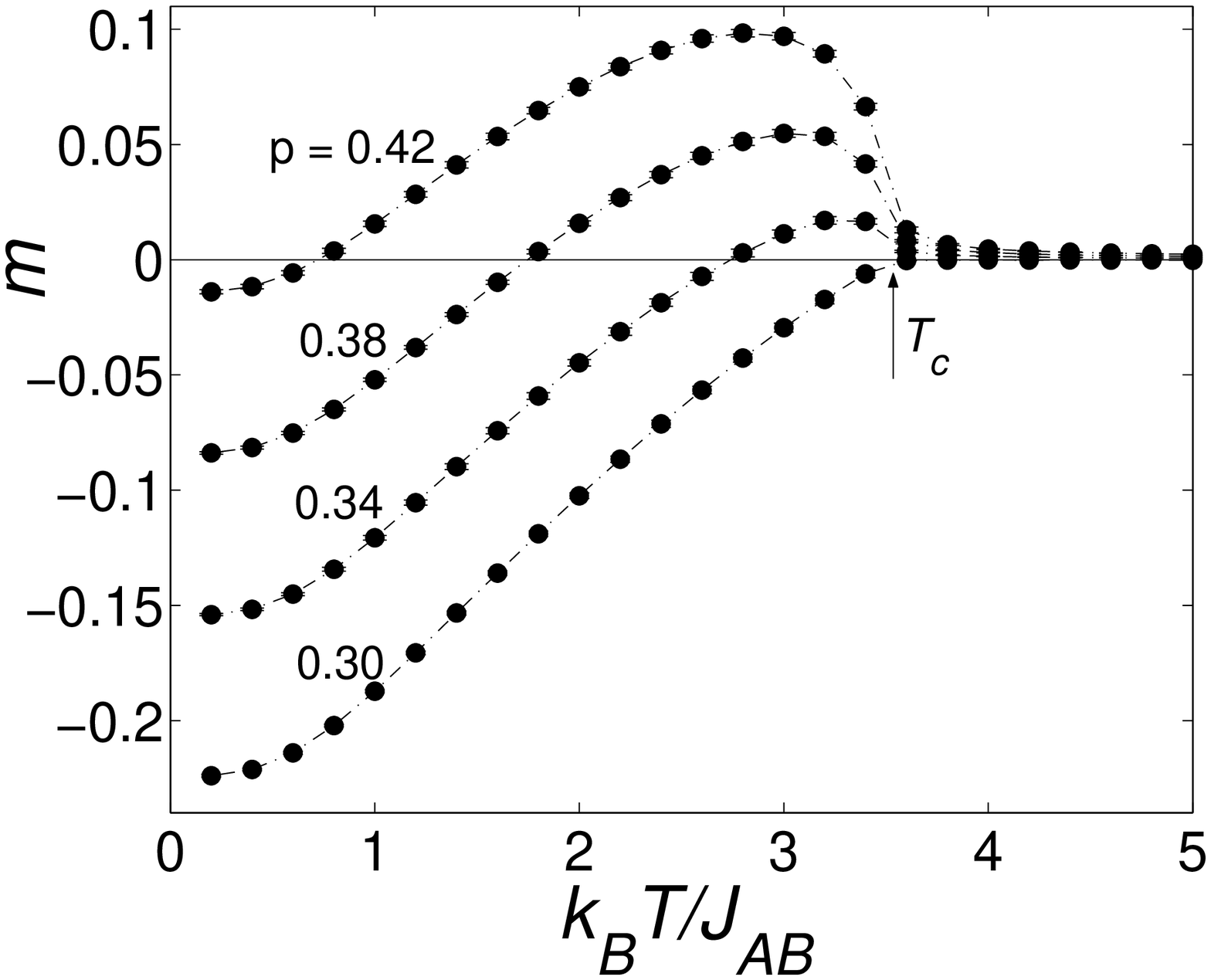}
\caption{Temperature dependences of the total magnetization $m$ for $R=0.45$ and different values of $p=0.42$, $0.38$, $0.34$, and $0.3$. The arrow approximately marks the position of the critical temperature.}
\label{fig:M-T}
\end{center}
\end{figure}
\begin{figure}[h]
\begin{center}
\includegraphics[scale=0.45]{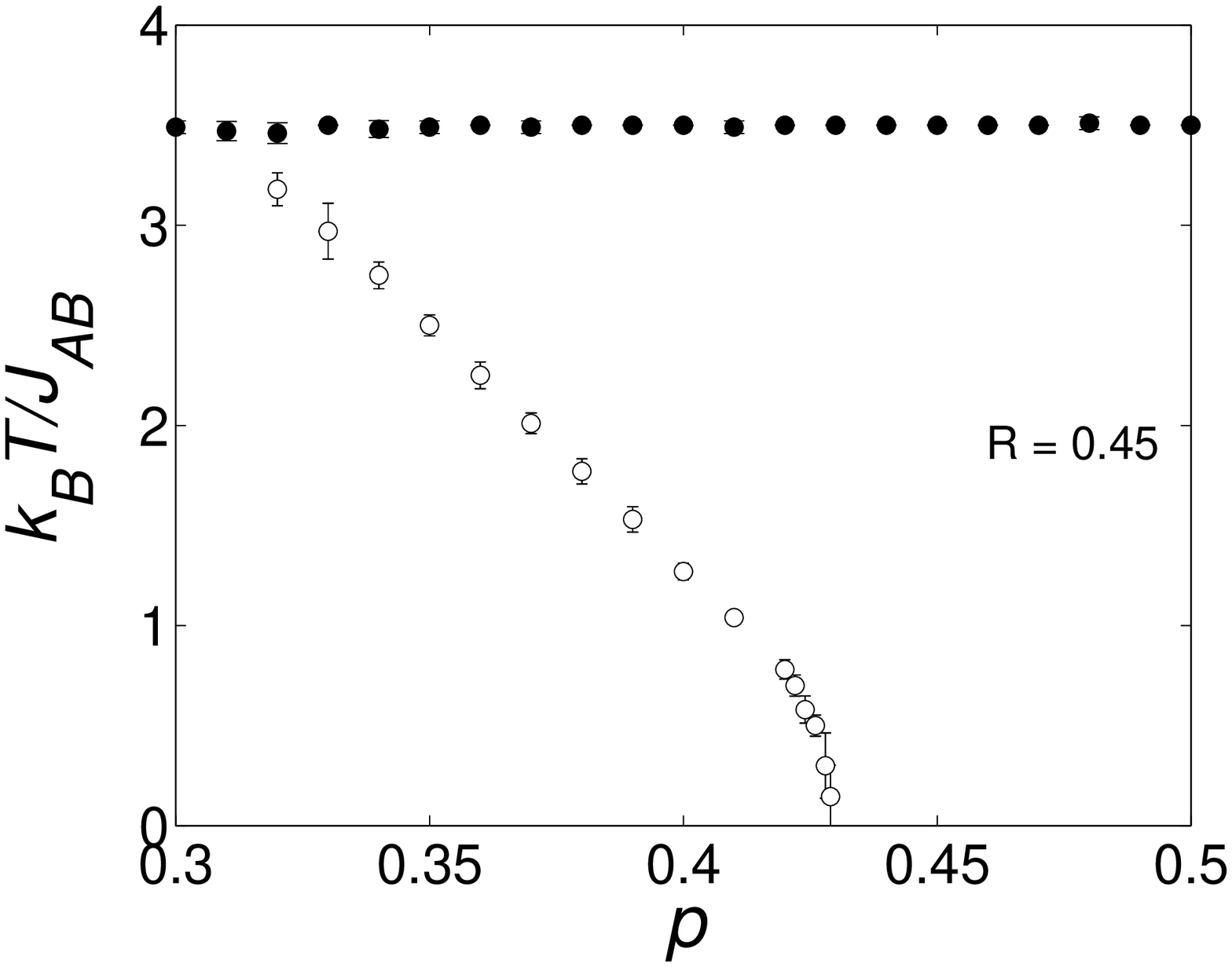}
\caption{Critical ($\bullet$) and compensation ($\circ$) temperatures as functions of $p$ for $R=0.45$.}
\label{fig:Tc_Tcom-p}
\end{center}
\end{figure}

The compensation temperature $T_k$ is the temperature at which total magnetization of the system $m$ vanishes below its critical temperature $T_c$ and it is determined by the condition $m(T_k)=0$. In this subsection we do not investigate in detail the compensation behaviour of the current model for various values of the parameters $R$ and $p$ but rather focus on the situation corresponding to the real Prussian blue analog $({\rm Ni}^{{\rm II}}_p{\rm Mn}^{{\rm II}}_{1-p})_{1.5}$ $[{\rm Cr}^{{\rm III}}({\rm C}{\rm N})_6]\cdot n {\rm H}_2 {\rm O}$ case, i.e. the case of $R=0.45$, for which we can confront our results with some experimental findings \cite{Ohko3,Ohko1}. For this value of the exchange ratio we found a compensation behaviour for the concentration values $p$ ranging approximately between $0.305$ and $0.425$, which roughly coincides with the results predicted by MFT \cite{Boba3} and EFT \cite{Hu1}. Some examples of the variation of the total magnetization with temperature in this concentration region are shown in Fig. \ref{fig:M-T}. The points at which the magnetization curves change sing below the respective critical temperatures represent the compensation temperatures. The variation of the critical (filled circles) and compensation (empty circles) temperatures with the concentration $p$ is presented in Fig. \ref{fig:Tc_Tcom-p}. These theoretical results are in line with the experimental magnetization measurements \cite{Ohko1}, in which the compensation behaviour was observed for the concentrations $0.38\leq p \leq 0.42$. The lower bound of $p$ was not explored in this experimental study, but the susceptibility measurements in \cite{Ohko3} suggest that the concentration at which the compensation line joins the critical one is close to $p=0.30$ i.e., close to the theoretical predictions.    

\begin{figure}[ht]
\begin{center}
\includegraphics[scale=0.45]{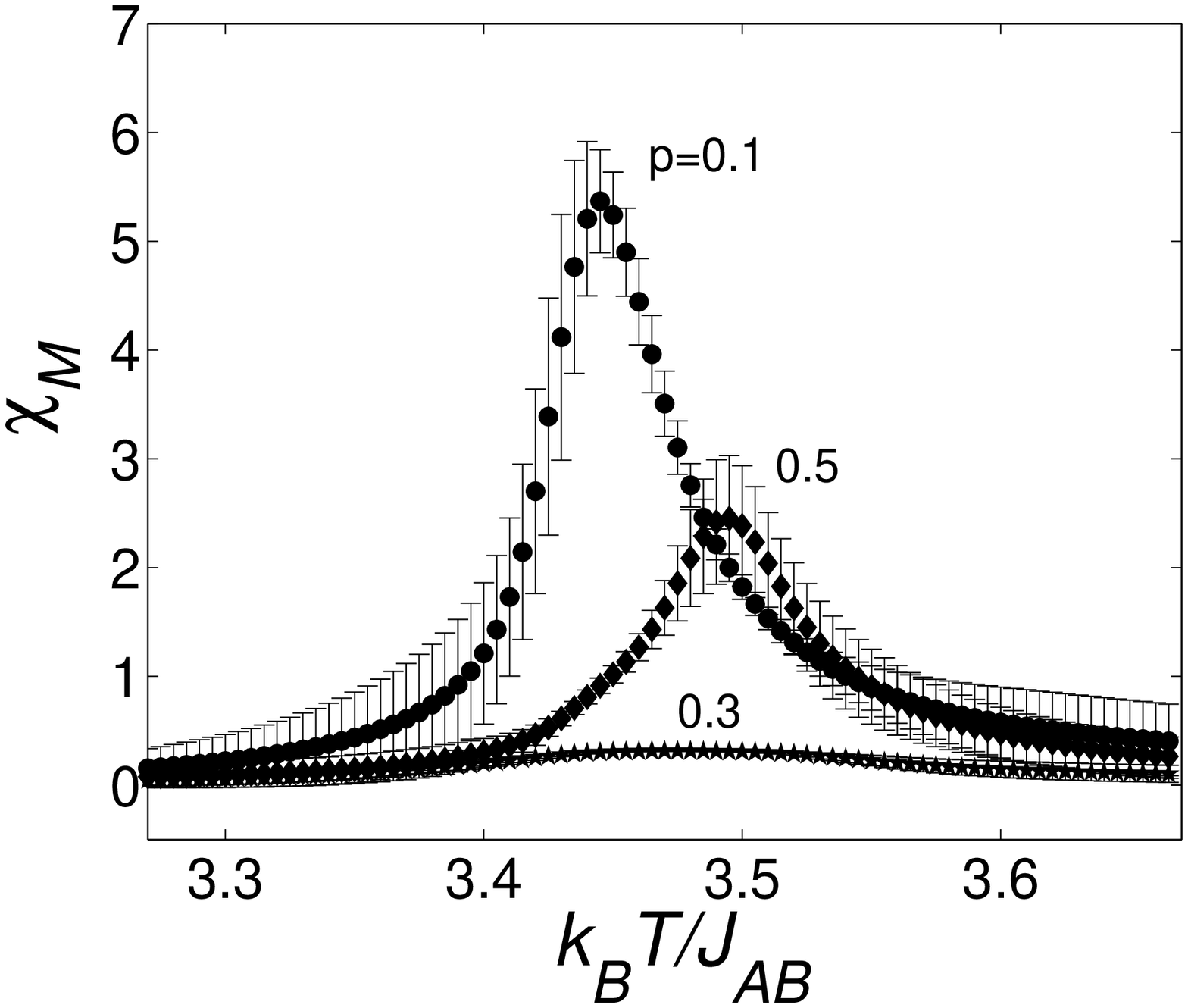}
\caption{Temperature dependences of the direct susceptibility $\chi_M$ for $R=0.45$, $L=32$, and $p=0.1$, $0.3$, and $0.5$.}
\label{fig:Xi-T}
\end{center}
\end{figure}

\begin{figure}[ht]
\begin{center}
\includegraphics[scale=0.45]{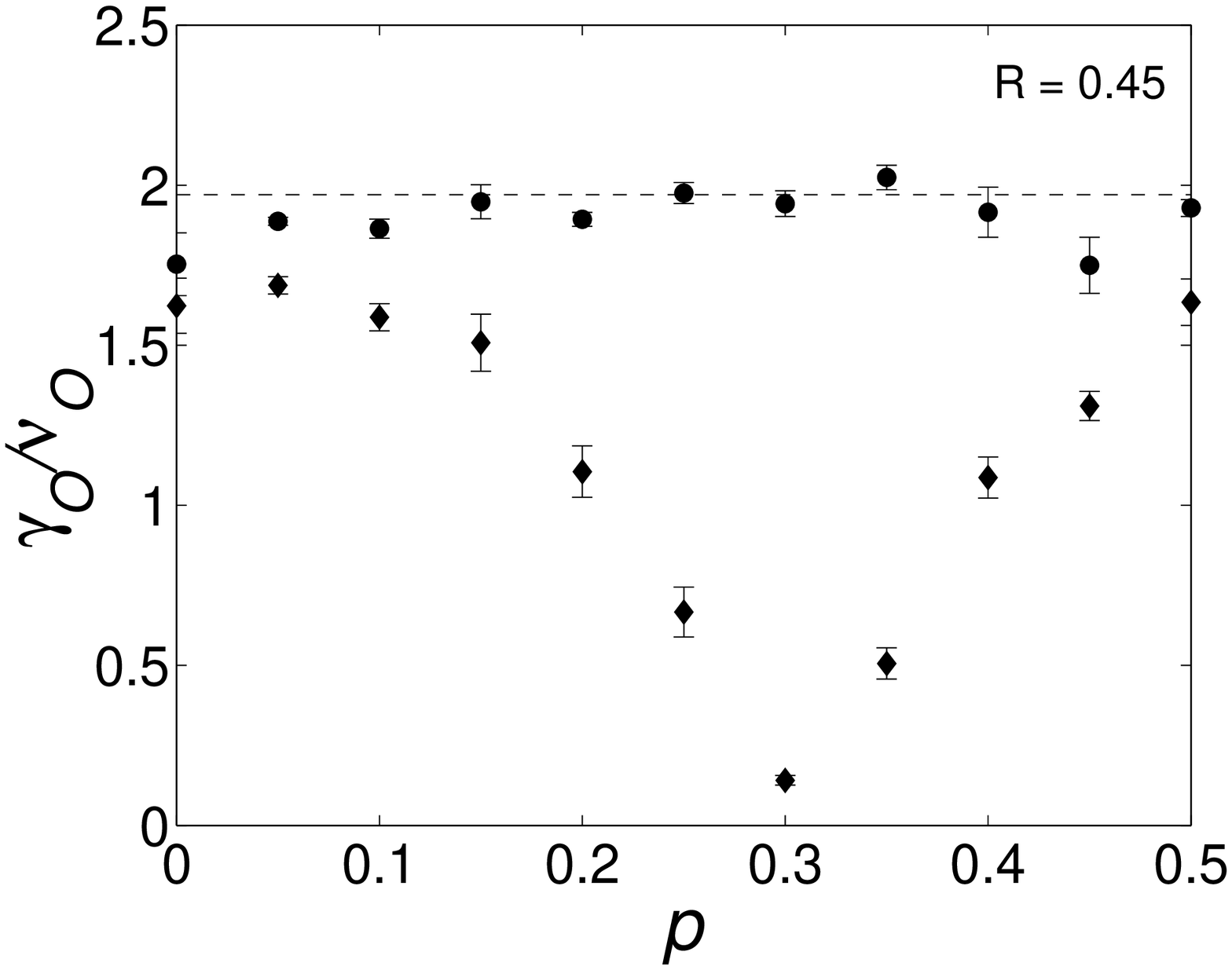}
\caption{Ratios of the critical exponents $\gamma_{M_s}/\nu_{M_s}$ (bullets) and $\gamma_{M}/\nu_{M}$ (diamonds), obtained from finite-size scaling  analyses of the staggered and direct susceptibilities, respectively, as functions of the mixing ratio $p$ for $R=0.45$.}
\label{fig:fss}
\end{center}
\end{figure}

The latter measurements revealed an unusual behaviour of the susceptibility at a critical temperature in the vicinity of the compensation point at $p=0.30$. Namely, for $p=0.30$ the paramagnetic susceptibility did not diverge near $T_c$. Motivated by this finding, we simulated the susceptibility for the same concentration and compared its behaviour with that further away from $p=0.30$. The results are shown in Fig. \ref{fig:Xi-T}. While the curves for $p=0.1$ and $0.5$ show prominent peaks at the transition (proper divergence cannot be observed due to finite-size effects), the one for $p=0.3$ is almost flat. We note that a significant contribution to the relatively large error bars at the susceptibility peaks for $p=0.1$ and $0.5$, besides the fluctuations of the peaks' heights, come also from the fluctuations of the peaks' positions obtained from different realizations. On the other hand, the principal source of the errors away from the peaks, where normally relatively small fluctuations are expected, is decreasing reliability of the reweighted values from the HMC method as we move away from the simulation temperature (usually chosen close to the peak's location \cite{Ferr1,Ferr2}). Nevertheless, the small error bars for $p=0.3$ in the vicinity of the critical temperature indicate absence of the characteristic peak in all realizations. 

This phenomenon can be explained by observing temperature dependences of the magnetization curves in this region of the concentration $p$. As shown in Fig. \ref{fig:M-T}, for $p>0.3$ the compensation effect occurs due to different temperature dependences of the sublattice magnetizations. Namely, as the temperature approaches the critical value $T_c$ from below, the positive sublattice magnetizations ($m_A$, $m_B$) prevail over the negative one ($m_C$), resulting in the finite (positive) value of the total magnetization. However, as $p$ approaches the value of $0.3$ (from either side), the positive and negative sublattice magnetizations tend to cancel out just below $T_c$ (see the case of $p=0.3$ in Fig. \ref{fig:M-T}). This situation is characteristic for an antiferromagnetic system which exhibits zero total magnetization and non-divergent behaviour of the direct susceptibility at critical (or Ne\'{e}l) temperature. 

In order to explore this behaviour in more detail, we focused on the behaviour of the susceptibility critical exponents in the area where the critical and compensation temperature lines merge. The exponents are calculated from the finite-size scaling given by equation (\ref{eq.scalchi}), using the histogram MC simulations. Fig. \ref{fig:fss} demonstrates that while the ratio of the exponents $\gamma_{M_s}/\nu_{M_s}$, corresponding to the staggered susceptibility, retains the values close to the standard Ising value $\gamma/\nu \approx 1.97$ \cite{Ferr3} in the whole concentration range, the ratio $\gamma_{M}/\nu_{M}$, corresponding to the direct susceptibility, clearly deviates from this value in a relatively broad range of $p$. Namely, it gradually decreases as the value of $p=0.3$ is approached from either side and it practically vanishes at $p \approx 0.3$. Thus, this MC result clearly confirms the experimental \cite{Ohko3} and MFT \cite{Boba3} predictions that the direct susceptibility of the Prussian blue analog $( {\rm Ni}^{\rm II}_p {\rm Mn}^{\rm II}_{1-p})_{1.5}[{\rm Cr^{III}(CN)}_6] \cdot n {\rm H_2O}$ for a special value of $p$, for which the compensation line terminates at the phase transition boundary, does not diverge at the critical temperature. Furthermore, it suggests that the susceptibility critical exponents of the model with a compensation behaviour might be affected by the proximity of the compensation point. This will be subject to further investigation. 

\section{Conclusions}

We employed standard and histogram Monte Carlo simulations to study the critical and compensation properties of a ternary ferro-ferrimagnetic spin alloy of the type $AB_pC_{1-p}$ on a cubic lattice, focusing on the case with the lattice stoichiometry and the parameters corresponding to the Prussian blue analog $( {\rm Ni}^{\rm II}_p {\rm Mn}^{\rm II}_{1-p})_{1.5}[{\rm Cr^{III}(CN)}_6] \cdot n {\rm H_2O}$. We confirmed the existence of the critical superexchange interaction ratio $R_c$ for which the critical temperature does not depend on the mixing ratio of the ions on $B$ and $C$ sublattices, which had been predicted within the MFA \cite{Boba1} as well as MC simulations on a square lattice with equivalent sublattices ($1:1$ stoichiometry) \cite{Buen1}. More importantly, we found that the value of $R_c \approx 0.47$ is virtually insensitive to the change in the stoichiometry and, therefore the lattice coordination number, and is close to that predicted by MFT on the same lattice \cite{Boba1} and that evaluated by MC simulations on the square lattice \cite{Buen1}. In contrast to these results, a recent MC study on the present spin system with equivalent sublattices gave quite different value of $R_c = 0.513$ \cite{Kiscam09}. However, we believe that this result is erroneous. Our suspicion is based not only on the deviating value of $R_c$, but also an unusually high value of the critical temperature for $R_c$ or equivalently for $p=1$. The latter corresponds to the binary mixture of the spins $S_A = \frac{3}{2}$ and $S_B = 1$, and its critical temperature had already been established within MFT \cite{Abub} and EFT with correlations \cite{Boba4} to be respectively roughly equal and lower than that obtained by the MC simulation in \cite{Kiscam09}. However, it is well known that effective field theories generally overestimate the value of the critical temperature in comparison with MC simulations. 

Finally, we showed that for $R=0$ the long-range order does not survive down to $p=0$, as predicted by MFT, and we roughly estimated the bounds for the percolation threshold $p_c$. Focusing on the case of $R=0.45$ and $3:2$ stoichiometry, corresponding to the real compound $( {\rm Ni}^{\rm II}_p {\rm Mn}^{\rm II}_{1-p})_{1.5}[{\rm Cr^{III}(CN)}_6] \cdot n {\rm H_2O}$, we reproduced the experimentally observed phenomenon of a non-divergent direct susceptibility at the critical temperature in the vicinity of a compensation point. This peculiar behaviour of the direct susceptibility was demonstrated by evaluation of the critical exponents in the finite-size scaling analysis. 

Based on the above considerations, it can be concluded that the current Monte Carlo study brought both quantitative and qualitative improvement over the results previously obtained within MFT. Nevertheless, despite its simplicity, MFT in many respects did a good job of predicting the critical and compensation behaviour of the ternary alloy considered in the study. This finding corroborates that resulting from the experimental study of the magnetic properties of the Prussian blue analogs \cite{Hash1}.

\section*{Acknowledgments}
This work was supported by the Scientific Grant Agency of Ministry of Education of Slovak Republic (Grant No. 1/0128/08).



\begin{thebibliography}{20}

\bibitem{Verd1} M. Verdaguer, G.S. Girolami, Magnetic Prussian Blue Analogs, in: J.S. Miller, M. Drillon (Eds.), Magnetism: Molecules to Materials V, Wiley-VCH Verlag GmbH Co. KGaA, Weinheim, 2005.
\bibitem{Hash1} K. Hashimoto, S. Ohkoshi, Phil. Trans. R. Soc. Lond. A 357 (1999) 2977.
\bibitem{Ohko3} S. Ohkoshi, K. Hashimoto, Phys. Rev. B 60 (1999) 12820.
\bibitem{Ohko1} S. Ohkoshi, T. Iyoda, A. Fujishima, K. Hashimoto, Phys. Rev. B 56 (1997) 642.
\bibitem{Ohko4} S. Ohkoshi, Y. Abe, A. Fujishima, K. Hashimoto, Phys. Rev. Lett. 82 (1999) 1285.
\bibitem{Boba1} A. Bob\'ak, F.O. Abubrig, T. Balcerzak, Phys. Rev. B 68 (2003) 224405.
\bibitem{Dely1} J. Dely, A. Bob\'ak, Physica B 388 (2007) 49.
\bibitem{Boba2} A. Bob\'ak, O.F. Abubrig, D. Horv\'ath, Physica A 312 (2002) 187.
\bibitem{Hu1} H. Hu, Z. Xin, W. Liu, Phys. Lett. A 357 (2006) 388.
\bibitem{Buen1} G.M. Buend\'ia, J.E. Villarroel, J. Magn. Magn. Mater. 310 (2007) e495.
\bibitem{Tsuj1} S. Tsuji, T. Kasama, T. Idogaki, J. Magn. Magn. Mater. 310 (2007) e471.
\bibitem{Dely2} J. Dely, A. Bob\'ak, D. Horv\'ath, Acta Phys. Pol. A 113 (2008) 461.
\bibitem{Devi1} B. Deviren, O. Canko, M. Keskin, J. Magn. Magn. Mater. 321 (2009) 1231.
\bibitem{Dely3} J. Dely, A. Bob\'ak A, M. \v{Z}ukovi\v{c}, Phys. Lett. A 373 (2009) 3197.
\bibitem{Ferr1} A.M. Ferrenberg, R.H. Swendsen, Phys. Rev. Lett. 61 (1988) 2635.
\bibitem{Ferr2} A.M. Ferrenberg, R.H. Swendsen, Phys. Rev. Lett. 63 (1989) 1195.
\bibitem{Ferr3} A.M. Ferrenberg, D.P. Landau  Phys. Rev. B 44 (1991) 5081.
\bibitem{Boba3} A. Bob\'ak, J. Dely, T. Balcerzak, Czech. J. Phys. 54 (2004) D523.
\bibitem{Kiscam09} E. Kis-Cam, E. Aydiner, J. Magn. Magn. Mater. doi:10.1016/j.jmmm.2009.10.029.
\bibitem{Abub} F.O. Abubrig, D. Horv\'ath, A. Bob\'ak, M. Ja\v{s}\v{c}ur, Physica A 296 (2001) 437.
\bibitem{Boba4} A. Bob\'{a}k, Physica A 258 (1998) 140.

\end{thebibliography}
\end{document}